\documentclass{PoS}

\title{The excited hadron spectrum in lattice QCD using a 
 new variance reduction method}

\ShortTitle{Excited spectrum in lattice QCD}

\author{\speaker{Colin Morningstar}, A.~Bell, C.Y.~Chen, D.~Lenkner, C.H.~Wong\\
        Dept.~of Physics, Carnegie Mellon University, 
        Pittsburgh, PA 15213, USA}

\author{J.~Bulava\\
        NIC, DESY, Platanenallee 6, D-15738, Zeuthen, Germany}

\author{J.~Foley\\
        Physics Department, 
        University of Utah, 
        Salt Lake City, UT 84112, USA}

\author{K.J.~Juge\\
        Dept.~of Physics, University of the Pacific, Stockton, CA 95211, USA}

\author{M.~Peardon\\
        School of Mathematics, Trinity College, Dublin 2, Ireland}

\abstract{Progress in calculating the spectrum of excited baryons and mesons 
 in lattice QCD is described.  Correlation matrices of sets of
 spatially-extended hadron operators have been studied and their effectiveness
 in facilitating the extraction of excited-state energies is demonstrated.  
 First applications of the stochastic LapH method, a new method of 
 estimating the low-lying effects of quark propagation, are presented.}

\FullConference{35th International Conference of High Energy Physics - ICHEP2010,\\
		July 22-28, 2010\\
		Paris France}

\begin{document}

This talk is a progress
report on our efforts to study the excited-state spectrum of QCD using the
Monte Carlo method: results from our process of selecting optimal single-hadron
operators are presented, and a new method of estimating slice-to-slice quark
propagation is outlined.

The use of operators whose correlation functions $C(t)$ attain their
asymptotic form as quickly as possible is crucial for reliably
extracting excited hadron masses.  
Since excited hadrons are expected to be large objects, 
the use of spatially extended operators is a key ingredient in
the operator design and implementation.  A more detailed discussion
of these issues can be found in Ref.~\cite{baryons2005A}.  First results
using unquenched configurations were reported in Ref.~\cite{spectrum2009}.

\begin{figure}
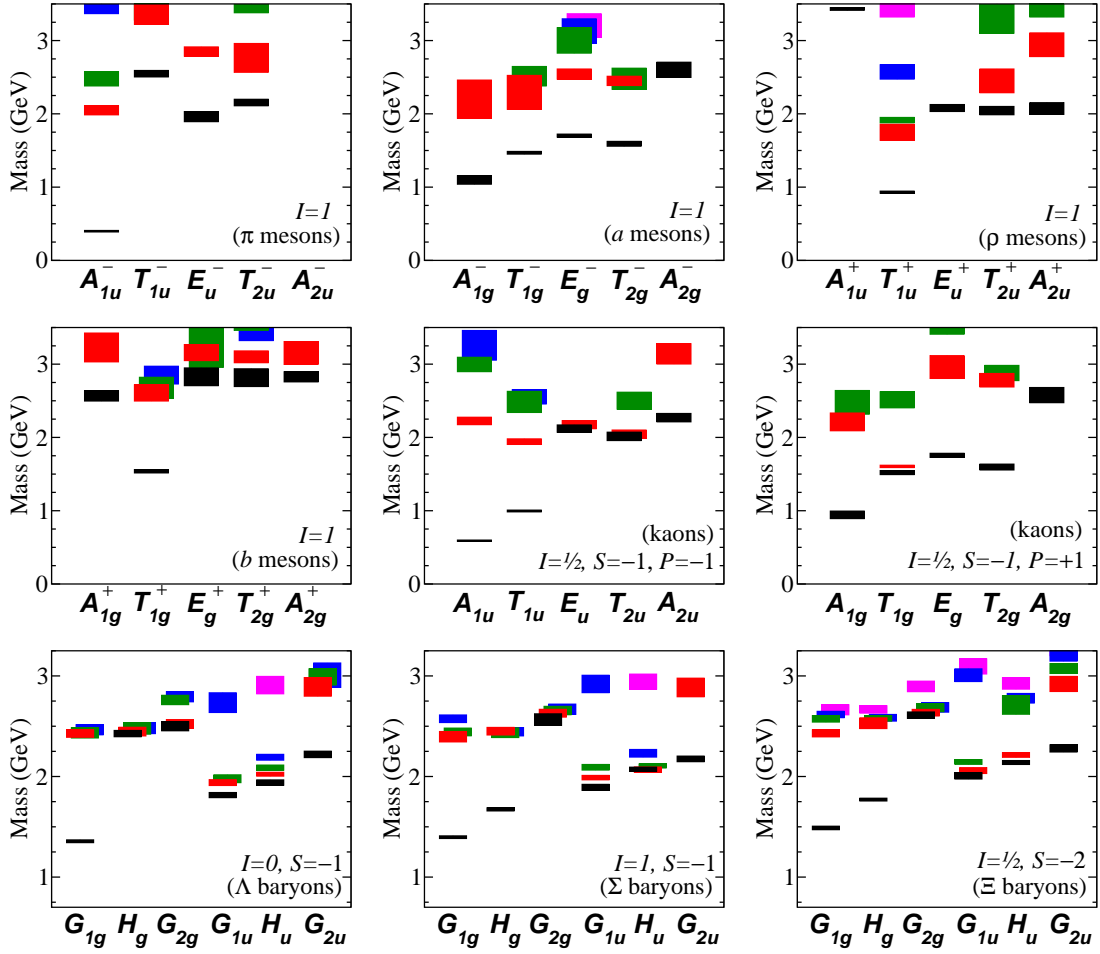

\begin{center}
\begin{minipage}{5.9in}
\includegraphics[width=1.8in,bb=31 34 523 461]{pi_mesons_pruning.eps}\quad
\includegraphics[width=1.8in,bb=31 34 523 461]{a_mesons_pruning.eps}\quad
\includegraphics[width=1.8in,bb=31 34 523 461]{rho_mesons_pruning.eps}\\[3mm]
\includegraphics[width=1.8in,bb=31 34 523 461]{b_mesons_pruning.eps}\quad
\includegraphics[width=1.8in,bb=31 34 523 461]{kaons_oddp_pruning.eps}\quad
\includegraphics[width=1.8in,bb=31 34 523 461]{kaons_evenp_pruning.eps}\\[3mm]
\includegraphics[width=1.8in,bb=31 34 523 461]{lambda_baryons_pruning.eps}\quad
\includegraphics[width=1.8in,bb=31 34 523 461]{sigma_baryons_pruning.eps}\quad
\includegraphics[width=1.8in,bb=31 34 523 461]{xi_baryons_pruning.eps}
\end{minipage}
\end{center}
\caption{
Hadron operator selection: low-statistics simulations have been performed
to study the hundreds of single-hadron operators produced by our group-theoretical
construction.  A ``pruning" procedure was followed in each channel to select
good sets of between six to a dozen operators.  The plots above show the
stationary-state energies extracted to date from correlation matrices of the finally
selected single-hadron operators.  Results were obtained using between 50 to 100 
configurations on a $16^3\times 128$ anisotropic lattice for $N_f=2+1$ quark flavors
with spacing $a_s\sim 0.12$~fm, $a_s/a_t\sim 3.5$, and quark masses such that
$m_\pi\sim 380$~MeV.  Each box indicates the energy of one stationary 
state; the vertical height of each box indicates the statistical error.
\label{fig:pruning}}
\end{figure}

A large effort has been undertaken to select optimal sets of baryon and meson
operators in a large variety of isospin sectors.
Low-statistics Monte Carlo computations were done to accomplish the
operator selections using between 50 to 100 configurations on a $16^3\times 128$ 
anisotropic lattice for $N_f=2+1$ quark flavors with spacing $a_s\sim 0.12$~fm, 
$a_s/a_t\sim 3.5$, and quark masses such that the pion has mass around 380~MeV. 
The method described in Ref.~\cite{distillation2009} was used.
Stationary-state energies using the finally selected operator sets are shown in
Fig.~\ref{fig:pruning}.  The nucleon, $\Delta$, $\Xi$, $\Sigma$, and $\Lambda$
baryons were studied, and light isovector and kaon mesons were investigated. 
Hundreds of operators were studied, and optimal sets containing eight or so
operators in each symmetry channel were found.  Future computations will focus
solely on the operators in the optimal sets.  See Ref.~\cite{wallace2010}
for our most recent high-statistics study.

To study a particular eigenstate of interest in the Monte Carlo method, all 
eigenstates lying below that state must first be extracted, and as the pion gets
lighter in lattice QCD simulations, more and more \textit{multi-hadron} states will
lie below the excited resonances.  In the evaluation of the temporal correlations 
of the multi-hadron operators that we intend to use, it is not possible to
completely remove all summations over the spatial sites on the source time-slice using
translation invariance.  Hence, the need for estimates of the quark propagators from 
all spatial sites on a time slice to all spatial sites on another time slice
cannot be sidestepped.  Some correlators will involve diagrams with
quark lines originating at the sink time $t$ and terminating at the
same sink time $t$, so quark propagators involving
a large number of starting times $t$ must also be handled.

A new way of stochastically estimating such slice-to-slice quark 
propagators was introduced in Ref.~\cite{hadron2009}.  The first ingredient in
the method is the use of a new Laplacian Heaviside (LapH) quark-field smearing
scheme defined by
$
\widetilde{\psi}(x) = 
 \Theta\left(\sigma_s^2+\widetilde{\Delta}\right)\psi(x),
$
where $\widetilde{\Delta}$ is the three-dimensional covariant Laplacian
in terms of the stout-smeared gauge field and $\sigma_s$ is the smearing cutoff
parameter.  The Heaviside function truncates the sum over Laplacian
eigenmodes, restricting the summation to some number $N_v$ of the lowest-lying
$-\widetilde{\Delta}$ eigenmodes on each time slice.  Quark propagation
is then estimated using $Z_N$ noise introduced in the LapH subspace. 
The noise vectors $\varrho$ have spin, time, and Laplacian eigenmode number
as their indices.  Variance reduction is achieved by \textit{diluting} the noise
vectors\cite{alltoall}. A given dilution scheme can be viewed as the application
of a complete set of projection operators.  Our dilution projectors are
products of time dilution, spin dilution, and Laph eigenvector dilution
projectors.  For each type (time, spin, Laph eigenvector) of dilution, we
studied four different dilution schemes: none, full, blocking, and interlacing.
We use a triplet (T, S, L) to specify a given dilution scheme, where ``T" denote 
time, ``S" denotes spin, and ``L" denotes Laph eigenvector dilution.  The schemes
are denoted by 1 for no dilution, F for full dilution, and B$K$ and I$K$ for
block-$K$ and interlace-$K$, respectively.  For example, full time and spin
dilution with interlace-8 Laph eigenvector dilution is denoted by
(TF, SF, LI8). The volume dependence of this new method was found to be very mild,
making the method suitable for large volume calculations.

\begin{figure}
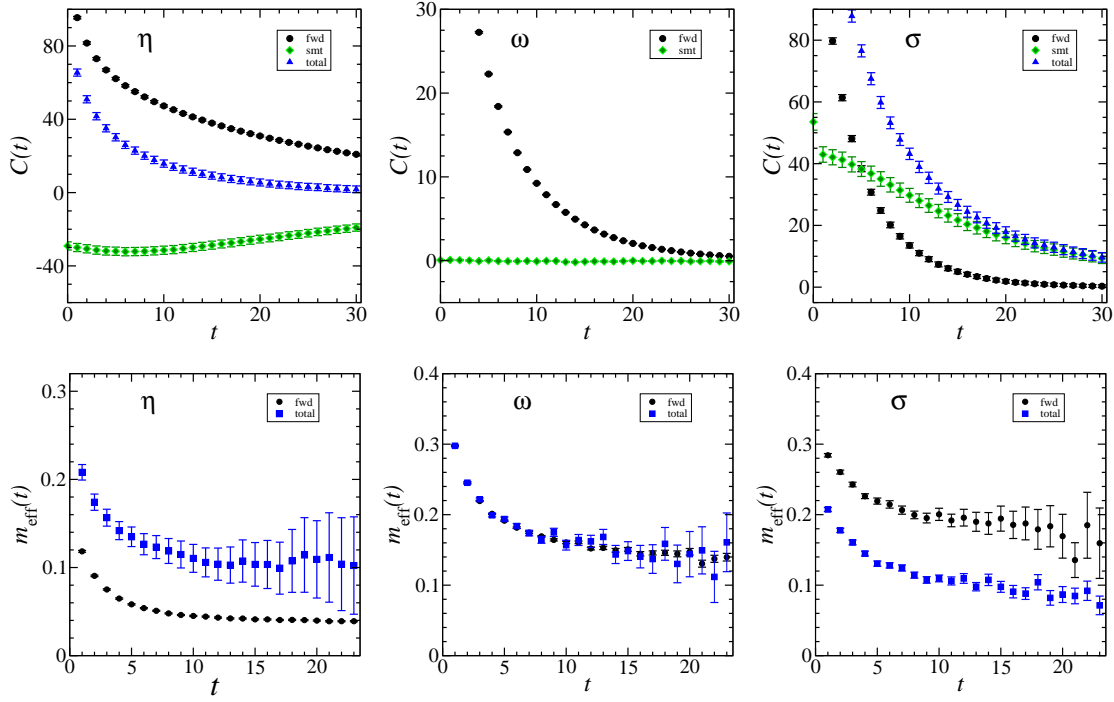

\begin{center}
\begin{minipage}{5.9in}
\includegraphics[width=1.8in,bb=21 34 523 530]{corr860.PS.eps}\quad
\includegraphics[width=1.8in,bb=21 34 523 530]{corr860.V.eps}\quad
\includegraphics[width=1.8in,bb=21 34 523 530]{corr860.S.eps}\\[3mm]
\includegraphics[width=1.8in,bb=21 34 543 530]{effmass860.PS.eps}\quad
\includegraphics[width=1.8in,bb=21 34 543 530]{effmass860.V.eps}\quad
\includegraphics[width=1.8in,bb=21 34 543 530]{effmass860.S.eps}\\[-7mm]
\end{minipage}
\end{center}
\caption{
Correlators $C(t)$ and effective masses $m_{\rm eff}(t)$ against
temporal separation $t$ for single-site operators which produce
the isoscalar pseudoscalar $\eta$, vector $\omega$, and scalar $\sigma$ mesons.
Results were obtained using 198 configurations with $N_f=2+1$ flavors of
quark loops on a $24^3\times 128$ anisotropic lattice with spacing
$a_s\sim 0.12$~fm and aspect ratio $a_s/a_t\sim 3.5$ for a pion mass
$m_\pi\sim 220$~MeV. In the legends, ``fwd" refers to contributions
from the diagram containing only forward-time source-to-sink quark lines,
``smt" refers to contributions from the diagram containing only
quark lines that originate and terminate at the same time.  For the 
$\sigma$ channel, the ``smt" contribution has a vacuum expectation value
subtraction. Forward-time quark lines use dilution scheme (TF, SF, LI8)
and same-time quark lines use (TI16, SF, LI8).
\label{fig:isoscalars860}}
\end{figure}

Results for three isoscalar mesons are shown in Fig.~\ref{fig:isoscalars860}.
Such mesons are notoriously difficult to study in lattice QCD, but the new
method appears to produce estimates of their temporal correlations with 
unprecedented accuracy. These plots suggest that evaluating correlation functions
involving our multi-hadron operators will be feasible with the stochastic
LapH method.

The next steps in our spectrum project are to combine our moving single-hadron
operators to form multi-hadron operators, then complete 
computations of QCD stationary-state energies using, for the first time, both 
single-hadron and multi-hadron operators.
This work was supported by the U.S.~National Science Foundation 
under awards PHY-0510020, PHY-0653315, PHY-0704171, and PHY-0969863 and 
through TeraGrid resources provided by the Pittsburgh Supercomputer Center, 
the Texas Advanced Computing Center, and the National Institute for Computational
Sciences under grant numbers TG-PHY100027 and TG-MCA075017.  MP is supported by 
Science Foundation Ireland under research grant 07/RFP/PHYF168.  The USQCD
QDP++/Chroma library\cite{chroma} was used in developing the software
for the calculations reported here. 

\bibliographystyle{aipprocl} 

\end{document}